# Observation of Distinct Bulk and Surface Chemical Environments in a Topological Insulator under Magnetic doping

*Ivana Vobornik$^{§*}$, Giancarlo Panaccione$^{§*}$, Jun Fujii$^{§}$, Zhi-Huai Zhu$^{ꞓ}$, Francesco Offi$^{&}$, Benjamin R. Salles$^{§}$, Francesco Borgatti$^{‡}$, Piero Torelli$^{§}$, Jean Pascal Rueff $^{‖¥}$, Denis Ceolin$^{‖}$, Alberto Artioli$^{§}$, Giorgio Levy$^{ꞓ¢}$, Massimiliano Marangolo$^{⊥}$, Mamhoud Eddrief$^{⊥}$, Damjan Krizmancic$^{§}$, Huiwen Ji$^{#}$, Andrea Damascelli$^{ꞓ¢}$, Gerrit van der Laan $^{†}$, Russell G. Egdell$^{∀}$ and Robert J. Cava$^{#}$*

$^{§}$Istituto Officina dei Materiali (IOM)-CNR, Laboratorio TASC, in Area Science Park, S.S.14, Km 163.5, I-34149 Trieste, Italy

$^{ꞓ}$Department of Physics & Astronomy, University of British Columbia, Vancouver, British Columbia V6T 1Z1, Canada

$^{†}$Dipartimento di Scienze, Università di Roma Tre, I-00146, Rome, Italy

‡ISMN-CNR, via Gobetti 101, I-40129 Bologna, Italy,

∥Synchrotron SOLEIL, L'Orme des Merisier, BP 48 Saint-Aubin, 91192 Gif sur Yvette, France,

¥Laboratoire de Chimie Physique – Matière et Rayonnement, Université Pierre et Marie Curie, CNRS, 11 rue Pierre et Marie Curie, 75005 Paris, France

⊥ Institut des NanoSciences de Paris, UPMC-CNRS UMR 7588, 4 place Jussieu, 75252 Paris Cedex 5, France

#Department of Chemistry, Princeton University, Princeton, New Jersey, 08544 USA

₡Quantum Matter Institute, University of British Columbia, Vancouver, British Columbia V6T 1Z4, Canada

†Diamond Light Source, Chilton, Didcot, Oxfordshire OX11 0DE, United Kingdom

∀Department of Chemistry, University of Oxford, Inorganic Chemistry Laboratory, South Parks Road, Oxford OX1 3QR, United Kingdom

**ABSTRACT** The influence of magnetic dopants on the electronic and chemical environments in topological insulators (TIs) is a key factor when considering possible spintronic applications based on topological surface state properties. Here we provide spectroscopic evidence for the presence of distinct chemical and electronic behavior for surface and bulk magnetic doping of $Bi_2Te_3$. The inclusion of Mn in the bulk of $Bi_2Te_3$ induces a genuine dilute ferromagnetic state, with reduction of the bulk band gap as the Mn content is increased. Deposition of Fe on the $Bi_2Te_3$ surface, on the other hand, favors the formation of iron telluride already at coverages as low as 0.07 monolayer, as a consequence of the reactivity of the Te-rich surface. Our results identify the factors that need to be controlled in the realization of magnetic nanosystems and interfaces based on TIs.

Topological Insulators (TIs) are promising materials for a new class of spin-based nanoelectronics with long spin coherence times and possible fault-tolerant information storage, due to the intrinsically 2-dimensional character of their surface states coupled with their protected spin environment[1-3]. A promising direction that goes beyond conventional methodologies is to tailor novel properties by exploiting TI-based heterostructures. This perspective has led to an intense research effort focused on magnetic doping, but attempts to control the magnetic properties of TIs have revealed a complex phenomenology. A stable ferromagnetic state has been obtained by bulk doping, via the inclusion of 3$d$-metal impurities (i.e. Mn, or Cr) in both single crystal and thin film forms of TIs[4-7] and deposition of a ferromagnetic layer onto the surface of a TI has resulted in a magnetic proximity effect[8-11], i.e. a magnetic coupling between the top ferromagnetic layer and the bulk dopant. However, a clear understanding of the modification of the electronic and chemical environments induced by the doping, and in particular at the surface/interface of the TIs has not been established. As for the case of diluted magnetic semiconductors, challenges remain[12,13]: for example, (i) tuning the dilution limit of the magnetic element in order to increase the ferromagnetic transition temperature to device-ready conditions, (ii) controlling the chemical environment upon adsorption or deposition onto the surface, and (iii) controlling the variation in the carrier concentration both at the surface and in the bulk. In diluted magnetic semiconductors, the stable ferromagnetic state is set via carrier mediated exchange, which depends on the carrier concentration and in turn on the magnetic dopant concentration[12,13]. The high density of free carriers required in such systems is unsuitable in the case of TIs, however, because low bulk conduction and high surface conduction are prerequisites for possible future TI-based spintronics devices.

The difference between surface and bulk properties is linked to a central issue in TI research: the role of dimensionality. The evidence for long range ferromagnetism in TIs is primarily based upon bulk sensitive techniques, and clear results distinguishing between surface and bulk properties are scarce. In TIs, a thickness dependent evolution of the spin and band structure exists[14,15]. As such, the critical thickness for the appearance of topological conditions (i.e. robustness against scattering, gapless surface states, etc) has been determined to be at least one quintuple layer in the $Bi_2X_3$ systems[14]. Moreover, it has been recently revealed that the surface state Dirac fermions are characterized by a layer-dependent entangled spin-orbital texture[15-17]. The coexistence of bulk ferromagnetism and topologically protected surface states has been clearly demonstrated, yet the influence of local magnetic impurities on the density of states is subject of an intense debate, mainly devoted to the opening of a gap in the Dirac cone[6].

Here we present the evolution of the chemical state and electronic properties in a prototypical TI, $Bi_2Te_3$, where we have taken both approaches to inducing magnetism: bulk magnetic doping, obtained by introducing Mn atoms in the pristine $Bi_2Te_3$ crystal structure, and surface magnetic doping, obtained by depositing a ferromagnetic Fe layer onto $Bi_2Te_3$. We find that the presence of Mn affects the electronic structure of bulk $Bi_2Te_3$, causing the reduction of the original bulk band gap. Hard X-ray Photoelectron Spectroscopy (HAXPES) and X-ray Absorption Spectroscopy (XAS), which provide bulk sensitive information with different information depths, confirm that the impurity-like electronic character of Mn is kept in the bulk, with no cluster formation, for Mn concentrations as high as x= 0.09. We further address the question of dilution and surface reactivity (i.e. whether 'surface doping' is possible) via deposition of magnetic elements on the surface of a TI (in our case Fe on $Bi_2Te_3$ and $Bi_{2-x}Mn_xTe_3$). STM analysis shows important modifications occurring at the surface of $Bi_{2-x}Mn_xTe_3$ upon Fe deposition: even for Fe thicknesses as low as 0.07 monolayer (0.1 Å), nm-wide structures co-exist with isolated Fe atoms; for coverages in excess of 0.35 monolayer (0.5 Å), larger structures predominate. Moreover, analysis of the core level spectra in soft X-ray PES provides clear evidence of chemical reaction between Te and Fe metal deposited on the surface, with reduction of Bi to a lower valence state and the concomitant modification of the Te electronic character. Simple thermochemical considerations lead us to point to the formation of an FeTe phase and either Bi metal or a subtelluride phase $Bi_2Te$ at the surface as a result of the deposition.

Figure 1a shows an STM topography image of the (001) surface of a $Bi_{2-x}Mn_xTe_3$ crystal with x= 0.09 taken after in situ room temperature cleavage. A sketch of the crystal structure is presented in panel 1b. In agreement with previous low temperature STM topography results, the substitutional Mn atoms are visible as shadows (inset in Fig. 1a ) in positions that correspond to those of the Bi atoms under the Te surface layer exposed by cleaving[4,10,11].

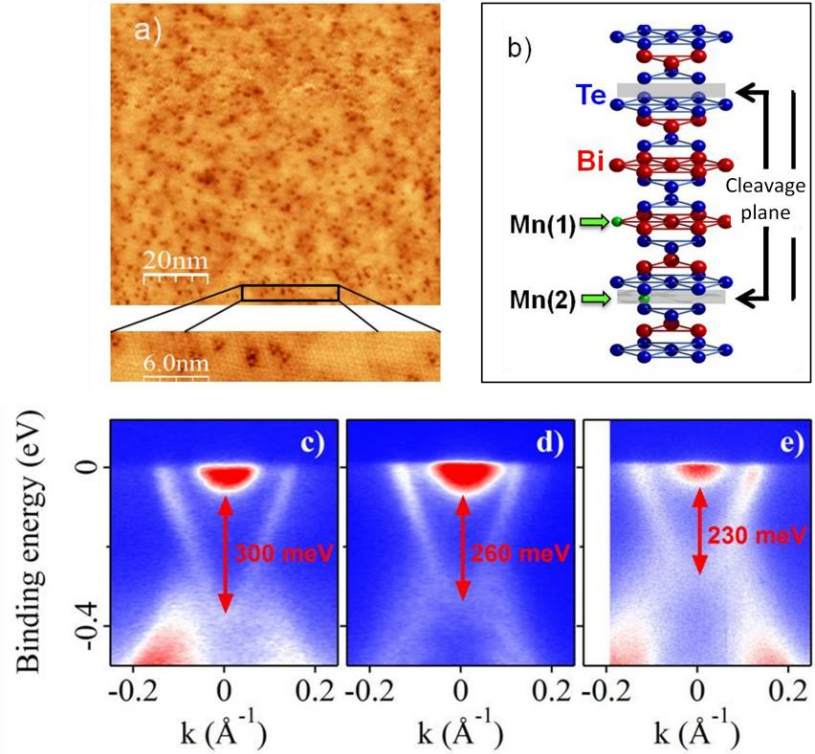

**Figure 1**. (a) STM topography image (100x100 nm$^2$) of an as-cleaved Bi$_{1.91}$Mn$_{0.09}$Te$_3$ (+500 mV, 0.18 nA) (001) surface. In the inset, the substitutional Mn atoms are visible as triangular shadows. (b) The crystal structure of Bi$_2$Te$_3$, with cleavage planes. Two sites for Mn atoms, Mn(1) and Mn (2), substitutional and interstitial respectively, are indicated by arrows. (c) Electronic band structure near E$_F$ measured by ARPES at T = 6 K. Dirac cone measured along Γ-K for (c) pristine Bi$_2$Te$_3$, and (d) and (e) Bi$_{2-x}$Mn$_x$Te$_3$ for x=0.04, and x= 0.09, respectively; the arrows give the measure of the bulk band gap, which progressively decreases as a function of Mn doping.

Quantitative analysis of the STM topography reveals that the number of Mn atoms at the surface is three times less than the nominal bulk doping, as has been observed previously[4]. This corresponds to a higher dilution of Mn in the surface region; part of the reduced doping value may be due to relaxation of the crystal structure after the cleavage procedure and consequent diffusion of Mn into a deeper layer[18]. Results of image analysis via cumulative probability of random pair separation (i.e. an estimate of the correlation between Mn pairs[4]), gives no indication of clustering effects. The evolution of the surface electronic structure near the Fermi level $E_F$ of $Bi_{2-x}Mn_xTe_3$ as measured by surface sensitive ARPES is presented in panels (c),(d) and (e) of Figure 1, for x= 0, 0.04, and 0.09 respectively. Since the Dirac cone in the as-cleaved Mn-doped samples is located above $E_F$ as a result of the hole doping due to Mn incorporation (see supporting information), we employed potassium deposition on the surface to rigidly shift the full band structure below $E_F$[19,20], thus allowing for the measurement of the bulk band gap at the Gamma point (i.e. the Valence Band-Conduction Band VB-CB energy difference measured at the Gamma point). Since the Dirac point coincides with the top of the bulk valence band $Bi_2Te_3$ (panel (a)), by assuming the rigid band shift we estimate the magnitude of the bulk energy gap as the energy difference between the Dirac point and the bottom of the bulk conduction band (the energy position of the Dirac point is extrapolated by assuming that the surface bands are linearly dispersing and cross each other at the Dirac point). As indicated by the arrows, we measure $\Delta E = 300 \pm 10$ meV in the pristine sample. We observe that the gap is reduced as the Mn content x increases, becoming $\Delta E = 260 \pm 10$ meV for x=0.04 and $\Delta E = 230 \pm 20$ for x= 0.09. The topological bulk band gap in $Bi_2Te_3$ is a direct consequence of band inversion, which is driven by the spin-orbit coupling (SOC) of the system[1]. The substitution of Bi atoms with Mn, which is an element with almost zero SOC, suppresses the band inversion by diluting the effective SOC in the solid, and consequently results in a reduced bulk band gap (and eventually in the disappearance of the topological properties altogether). In addition, the Mn impurity-induced suppression of band inversion could be further enhanced by near Fermi-level impurity states, as reported for $(Bi_{1-x} In_x)_2Se_3$ [21]. Unfortunately, the fact that the Dirac point of the topological surface state in $Bi_2Te_3$ is buried inside of the bulk valence-band manifold, makes it difficult to determine whether the Mn impurities open a gap at the Dirac point below the ferromagnetic transition temperature.

The impurity-like character of the Mn-atoms in the $Bi_2Te_3$ matrix is confirmed by the spectroscopic results presented in Figure 2. In panel (a) the Mn $L_{2,3}$ XAS spectra from $Bi_{1.91}Mn_{0.09}Te_3$ are compared to the reference Mn ones from the diluted magnetic semiconductor (Ga,Mn)As (Mn = 5 %) and from a metallic MnAs film. The dilute Mn systems display a broad double peak structure at the $L_2$ edge and smooth shoulders near the $L_3$ edge, corresponding to a hybrid $d^4$-$d^5$-$d^6$ configuration, with dominant $d^5$ character, as already discussed in previous reports[22-24]. These spectroscopic features are clearly visible in the Mn spectrum from $Bi_{1.91}Mn_{0.09}Te_3$. On the contrary, Mn in the metallic environment found in MnAs displays only a single component in the $L_2$ region, a different overall line shape, and a different $L_3/L_2$ ratio[22,23]. Knowing that the XAS signal averages over a total information depth of ~5 nm, while the probing depth for Hard X-ray PES (HAXPES) is >15nm[25,26], we obtain depth sensitive information shown in panel (b) of Figure 2; the Mn $1s$ core lines (~ 6540 eV of binding energy) from $Bi_{1.91}Mn_{0.09}Te_3$ are presented, measured with 11 keV of photon energy. The use of high photon energy guarantees truly bulk sensitive information, while maintaining chemical sensitivity and no overlap with lines from Bi and Te.

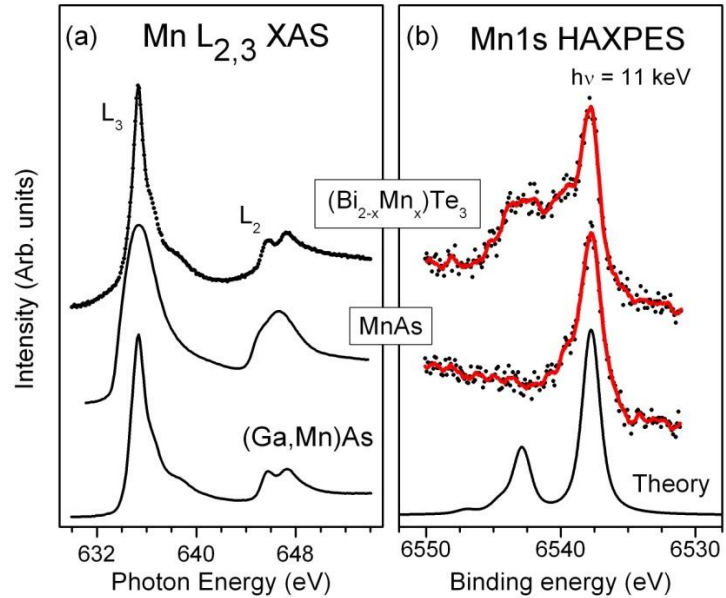

**Figure 2** (a) The normalized Mn $L_{2,3}$ XAS spectra (black circles) from a $Bi_{1.91}Mn_{0.09}Te_3$ sample, a reference 250 nm metallic MnAs film grown on GaAs and a reference (GaMn)As (5% Mn) sample. (b) The Mn 1s HAXPES spectra measured at $hv$ = 11 keV (black points; the red curves are guides to the eye.) from a $Bi_{1.91}Mn_{0.09}Te_3$ sample and from a reference 250 nm metallic MnAs film grown on GaAs compared to the calculation (black line). For the sake of comparison, the spectra have been aligned and normalized to the main line energy position and intensity. No background subtraction was applied. Spectra from $Bi_{1.91}Mn_{0.09}Te_3$ display a double structure as opposed to the single peak of the metallic sample. PES Calculations were performed using the parameters $\Delta = E(d^6) - E(d^5) = 1$ eV, $Q(1s,3d) = 3.75$ eV, $U(3d,3d) = 2.75$ eV, and a hybridization parameter $V = 2.5$ eV. This give gives a satellite peak at 5.2 eV above the main peak with 30% of the total intensity. All XAS and HAXPES spectra have been measured at room temperature and under identical experimental conditions.

Again, we compare spectra from $Bi_{1.91}Mn_{0.09}Te_3$ to those of the MnAs reference. The analysis of the HAXPES spectra not only confirms a distinct difference between the doped Mn in $Bi_{1.91}Mn_{0.09}Te_3$ and the Mn with metallic character in MnAs, but also reveals the presence of a double peak structure. To understand this structure, the photoemission for the Mn transition $3d^n \rightarrow 1s^1 3d^n \varepsilon$ can be calculated using an Anderson impurity model and taking into account the configuration interaction for the initial and final states[27]. For Mn core-level photoemission it is sufficient to consider only the $d^5$ and $d^6$ states, following previous results obtained for Mn $2p$ photoemission from (Ga,Mn)As [28]. An important difference between a $1s$ and $2p$ core hole is the degree of localization as well as screening, which affects the $3d$ valence states. The core-valence Coulomb interaction, $Q(1s,3d)$, and the intra-atomic Coulomb interaction, $U(3d,3d)$, will be smaller in the presence of a $1s$ core hole than for a $2p$ hole[29]. However, $Q(1s,3d)$ is still large enough to pull down the $1s^1 3d^6$ state in energy below the $1s^1 3d^5$ state. Such a level reversal leads to the appearance of a strong satellite feature[30], as visible both in experiment and theory in fig. 2(b), and totally absent in the metallic phase represented by the MnAs spectrum. Our data cannot exclude the presence of Mn atoms located in the Van der Waals gap; however, a large quantity of atoms confined in the Van der Waals gap of a layered material would imply a significant change in size of the crystal structure, which is not reported in similar doped TIs[31,32]. Moreover, the absence of metallic-like features in the spectroscopic results of Figure 2 excludes a significant clustering effect of Mn, either at the surface or in the bulk, consistent with the STM results in Fig. 1a. It is important to underline that Mn displays favorable characteristics as a magnetic impurity in bulk TIs: while the present observation agrees well with previous reports supporting intrinsic diluted ferromagnetism in $Bi_{2-x}Mn_xTe_3$[4,9] and in Cr-doped $(Bi,Sb)_2Te_3$[33], in other cases (e.g. Fe in $Bi_2Se_3$), experimental reports have shown a strong preference to form clusters with a dilution limit close to zero and/or clear phase separation of the ferromagnetic phase from the TI[8,34,35]. These observations hold important consequences on the choice of the material for spintronics applications.

Knowing that a further relevant parameter towards the realization of TI-based devices is the surface reactivity, we now extend our analysis to surface magnetic doping, in our case the growth of Fe on the surface of $Bi_{1.91}Mn_{0.09}Te_3$ and $Bi_2Te_3$. It has been found that low coverages (up to 0.5% of a monolayer) of ferromagnetic elements, e.g. Fe and Co, do not break the time-reversal symmetry in the TI and do not open a gap at the Dirac point, as opposed to what is expected for bulk doping [6,36]. In Figure 3, STM results show the modification induced by 0.07 and 0.35 monolayer Fe deposition on $Bi_{1.91}Mn_{0.09}Te_3$, in panel a and b, respectively. From the analysis of the diameter and height of the Fe-induced structures, and keeping in mind that a variation of the apparent height/diameter corresponds to a variation of the electronic density, and is bias dependent, we identify a majority of smaller isolated structures (Fe(1) in the figure) with apparent diameter < 1 nm and height < 0.2 nm, and a minority of relatively large islands (Fe(2) in the figure) with diameters and heights up to 2 nm and 6 nm, respectively.

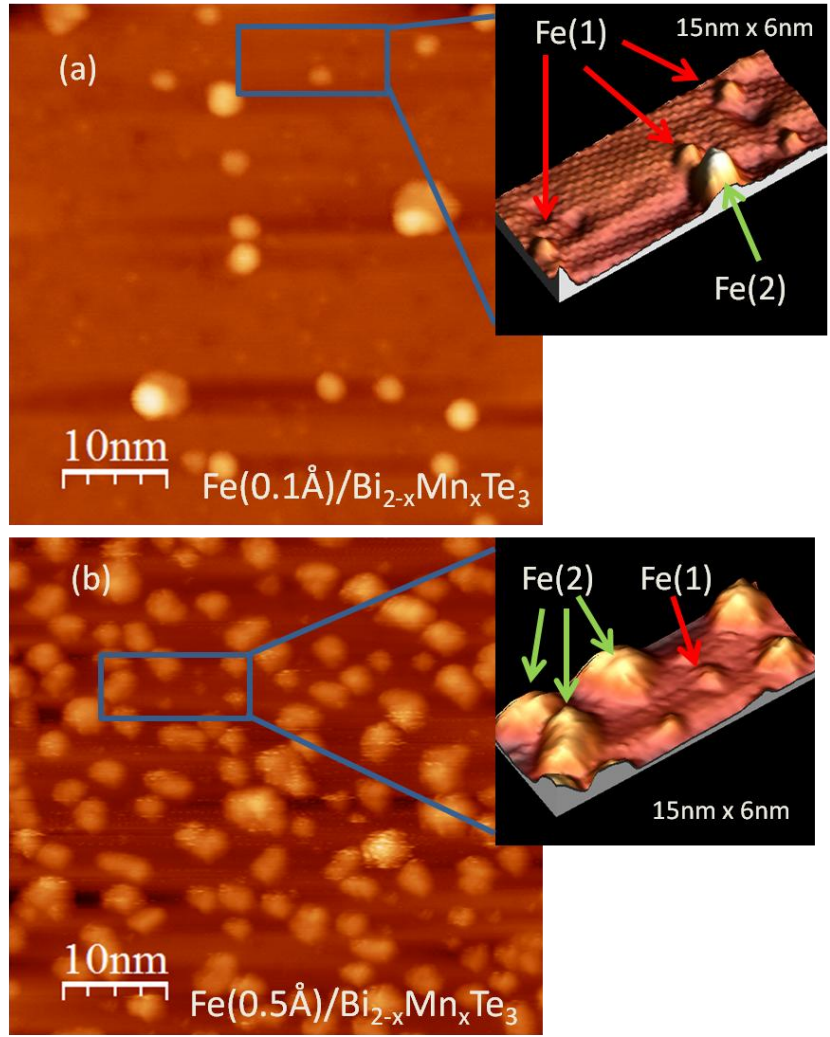

**Fig. 3** STM topography. (a) 0.1 Å (0.07 monolayer) Fe deposited on $Bi_{1.91}Mn_{0.09}Te_3$: left: inset: 3D rendering close-up (15 nm x 6 $nm^2$, +19 mV, 1.8 nA). Dark areas correspond to Mn atoms. Two kinds of Fe atoms, (Fe(1) and Fe(2), indicated by arrows) are found, corresponding to low and high apparent height. (b) 0.5 Å Fe (0.35 monolayer) on $Bi_{1.91}Mn_{0.09}Te_3$ (inset) 3D rendering close-up (15 nm x 6 $nm^2$, +680 mV, 0.29 nA). Isolated Fe(1)-type structures are rarely found while the density of large islands is high.

This observation, in agreement with similar results obtained on $Bi_2Se_3$ [11,37,38], points to the presence of two distinct kinds of Fe-related structures, where adatoms self-aggregate in cluster-like structures, while single impurities diffuse into subsurface layers. Increasing the Fe thickness to 0.35 monolayer not only produces the almost ubiquitous presence of large islands, with rare single structures, but also a relative uniformity of islands, as both height and diameter converge to a single average value (0.5±0.01 nm in height and 2 nm of diameter). These results suggest that a stable chemical and electronic condition has already been established at very low coverage. It is important to underline that the STM images in this case do not directly resolve the Fe atoms, as opposed to the case of Mn in the topography of $Bi_{2-x}Mn_xTe_3$ in Fig.1, but rather are the result of host states perturbed by the Fe[11]. At larger coverages, 4 monolayers and above, the islands are connected, the Fe is ferromagnetic, and the Mn magnetic bulk state found below 12 K in $Bi_{2-x}Mn_xTe_3$ can be raised to higher temperatures by the magnetic proximity effect produced by the Fe overlayer[9].

We obtain further information on the surface electronic and chemical structure of $Bi_2Te_3$ and $Bi_{2-x}Mn_xTe_3$ via the analysis of core level PES spectra, presented in Figure 4. Inspection of the core level spectra shows that the chemical environment of the $Bi_2Te_3$ host is not modified by the presence of Mn, with both Bi *5d* and Te *4d* core levels (panel b and c) substantially unchanged upon bulk doping. The analysis by a fitting procedure, the comparison with bulk sensitive PES spectra (see supporting information), and the ARPES data in Figure 1 agree well with the scenario that $Bi_2Te_3$ retains its topological properties upon the substitution of Mn[4,15]. Significant modifications are found, however, after Fe is deposited on the surface of $Bi_2Te_3$. These are shown in Figure 4 panel (a) and the close-up of the Bi *5d* and Te *4d* regions in panels (b) and (c). This suggests a profound change in the chemical state of the topmost layers of the TI, similarly to the Fe/$Bi_2Se_3$ and Co/$Bi_2Se_3$ cases in the low thickness (< 1 Angstrom) regime[10,39]. At a Fe coverage of 6 Å (i.e. a thickness able to sustain room temperature ferromagnetism), new components shifted to 0.7 eV *lower* binding energy appear in the Bi *4f* and Bi *5d* core level spectra, while in the Te *4d* region shows shifted and broadened components at 0.45 eV *higher* binding energy. This points to the presence of a chemical reaction between the deposited Fe and the $Bi_2Te_3$ that involves the oxidation of Te from the formal $Te^{2-}$ valence state and a concomitant reduction of the formal Bi valence state.

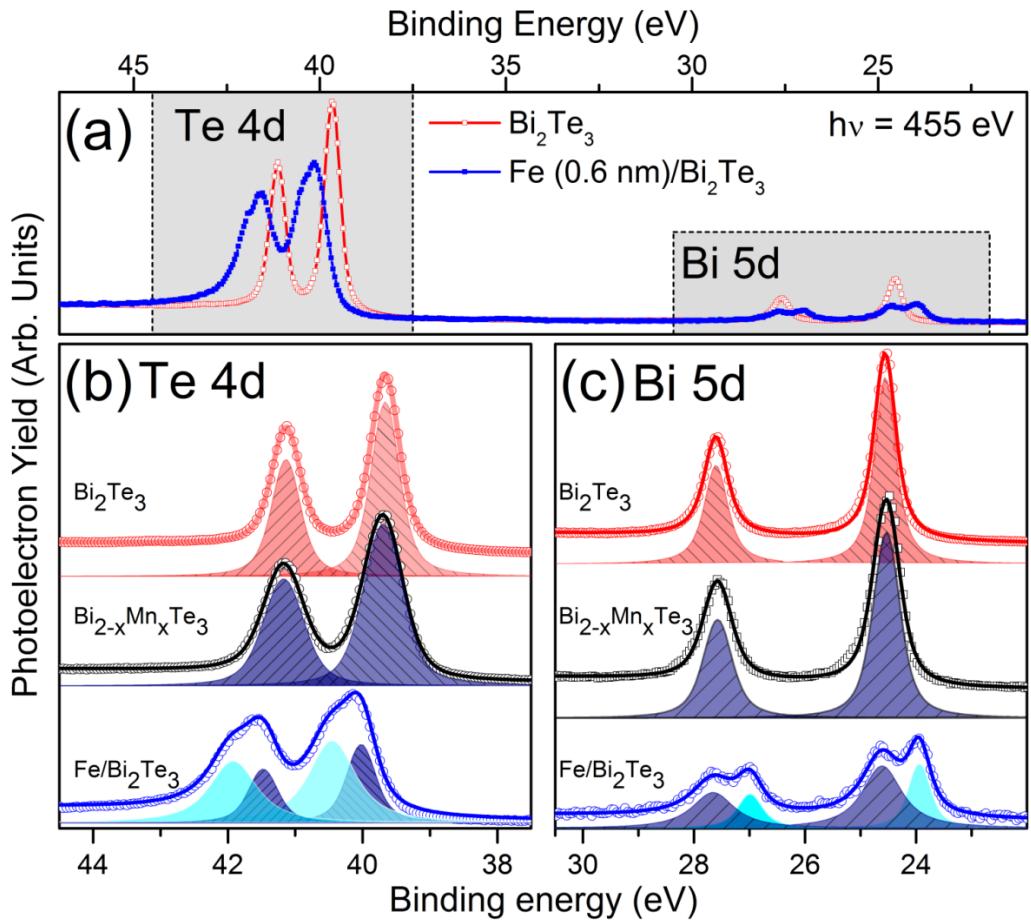

**Figure 4** (a) Soft x-ray photoemission (hν = 455 eV) from Te 4*d* and Bi 5*d* (symbols), measured in identical experimental conditions, before and after 6 Å Fe deposition on $Bi_2Te_3$. Close-up of the spectral region for Te 4*d* (b) and Bi 5*d* (c), showing fitting results (solid lines, superposed); the fitted Voigt functions for single peaks are shown as filled colored areas. In each panel, a comparison between pure $Bi_2Te_3$, $Bi_{1.91}Mn_{0.09}Te_3$ and Fe deposited on $Bi_2Te_3$ is presented. Spectra have been shifted on a vertical scale and normalized to the same area for sake of comparison. A Gaussian broadening of 0.25 eV has been included in the fitting procedure, to account for experimental energy resolution; a Shirley background has been subtracted from the raw data.

In addition, after quantitative analysis and normalization of the spectra, we note that the Bi core levels display a stronger attenuation than the Te core levels. This suggests that the Fe is mainly accommodated at the surface. Knowing that after cleave the exposed layer is Te-rich (see also Figure 1b), this observation agrees well with the STM results of Figure 3: the number of isolated Fe atoms, buried or intercalated, has already significantly decreased at 0.5 Å of Fe thickness, favoring a surface chemical environment containing both Fe and Te. We have explored the energetics of possible reactions between Fe and $Bi_2Te_3$ using standard thermodynamic tabulations[40]. We find that reactions involving the formation of $FeTe_2$ along with Bi metal or one of the Bi sub-tellurides (e.g. $Bi_2Te$ or BiTe) all involve negative free energies (G):

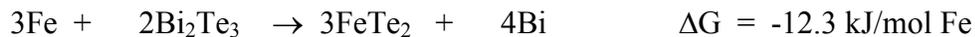
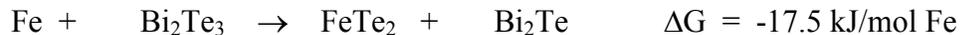
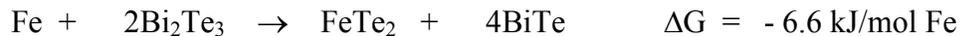

$3Fe + 2Bi_2Te_3 \rightarrow 3FeTe_2 + 4Bi \qquad \Delta G = -12.3 \text{ kJ/mol Fe}$

$Fe + Bi_2Te_3 \rightarrow FeTe_2 + Bi_2Te \qquad \Delta G = -17.5 \text{ kJ/mol Fe}$

$Fe + 2Bi_2Te_3 \rightarrow FeTe_2 + 4BiTe \qquad \Delta G = -6.6 \text{ kJ/mol Fe}$

In these reactions, the formal oxidation state of Te changes from -2 in $Bi_2Te_3$ to -1 in $FeTe_2$, i.e. a reduced partial charge on Te in $FeTe_2$ giving a shift to higher binding energy, while Bi is reduced from the +3 state to 0, +1 and +2 states respectively, giving rise to a shift in binding energy in the opposite direction. Thermodynamically, reduction to $Bi_2Te$ is most strongly favored, although all three reactions would account qualitatively for the directions of the observed chemical shifts. We note that these reactions are all favored with respect to formation of the iron rich FeTe phase, where thermochemical data[40] is given for "$FeTe_{0.9}$"[41].

$3Fe + 0.9Bi_2Te_3 \rightarrow 3FeTe_{0.9} + 1.8Bi \qquad \Delta G = -2.1$ kJ/mol Fe

$2Fe + 0.9Bi_2Te_3 \rightarrow 2FeTe_{0.9} + 0.9Bi_2Te \qquad \Delta G = -4.5$ kJ/mol Fe

$Fe + 0.9Bi_2Te_3 \rightarrow FeTe_{0.9} + 1.8BiTe \qquad \Delta G = +0.5$ kJ/mol Fe

A phase $Fe_7Te_8$ is also known but there appears to be no reliable thermochemical data for this phase. However it seems safe to assume that reaction free energies will be intermediate between those involved for formation of $FeTe_2$ and $FeTe_{0.9}$.

In summary, we have gained chemical insights into the electronic properties of $Bi_2Te_3$ under magnetic doping; we observe a profound difference between bulk and surface magnetic doping: the deposition of a ferromagnetic element, Fe, on the surface, produces in the near-surface region an important change of the surface chemistry, while Mn-bulk doping results in a homogeneous bulk phase, very much like conventional diluted magnetic semiconductors. Our results may help to identify the factors that need to be controlled in the realization of magnetic interface systems based on TIs: knowing that bulk ferromagnetic order may coexist with surface topological order, both the spin texture and the spin dynamics can be modified by controlled surface doping, and in general by thin film growth. The interaction between magnetic bulk textures and the charge of topologically protected surface states may help the realization of quantized anomalous Hall effect and magneto-electric coupling based applications.

ASSOCIATED CONTENT

**Supporting Information** contains details on calculation and sample preparation, additional experimental results (HAXPES, XPS and ARPES). This material is available free of charge *via* the Internet at http://pubs.acs.org.

AUTHOR INFORMATION

**Corresponding Authors**

*E.mail: vobornik@elettra.eu, vobornik@iom.cnr.it; panaccioneg@elettra.eu

**Author Contributions**

The manuscript was written through contributions of all authors. All authors have given approval to the final version of the manuscript. G.P.[§] and I.V.[§] contributed equally to the work.

ACKNOWLEDGMENT

Part of this research has been supported by the PIK project 'Ultraspin', funded by Sincrotrone Trieste. The authors are grateful to the teams of GALAXIES and APE beamlines for their support. The work at Princeton was supported by the US National Science Foundation grant DMR-0819860. The work at UBC was supported by the Max Planck - UBC Centre for Quantum Materials, the Killam, Alfred P. Sloan, and NSERC's Steacie Memorial Fellowships (A.D.), the Canada Research Chairs Program (A.D.), NSERC, CFI, and CIFAR Quantum Materials. This work has been partially funded by the Peter Wall Institute for Advanced Studies (UBC) under the international research scholars programme 2011-2013.

**(18)** Both the procedure of counting Mn atoms and the cumulative probability have been performed on several topographies from randomly chosen areas of the surface, on three different samples freshly cleaved, yielding similar results.

**(19)** Zhu, Z.-H.; Levy, G.; Ludbrook, B.; Veenstra, C.N.; Rosen, J.A.; Comin, R.; Wong, D.; Dosanjh, P.; Ubaldini, A.; Syers, P.; et al. Rashba spin-splitting control at the surface of the topological insulator $Bi_2Se_3$, *Phys. Rev. Lett.* **2011**, *107*, 186405.

**(20)** Hsieh, D.; Xia, Y.; Qian, D.; Wray, L.; Meier, F.; Dil, J.H.; Osterwalder, J.; Patthey, L.; Fedorov, A.V.; Lin, H.; et al. Observation of time-reversal-protected single-Dirac-cone topological insulator states in $Bi_2Te_3$ and $Sb_2Te_3$. *Phys. Rev. Lett.* **2009**, *103*, 146401.

**(21)** Wu, L.; Brahlek, M.; Valdés Aguilar, L.; Stier,A.V. ; Morris, C.M.; Lubashevsky, Y.; Bilbro,L.S.; Bansal, N.; Oh, S.; Armitage, N.P.;  A sudden collapse in the transport lifetime across the topological phase transition in $(Bi_{1-x}In_x)_2Se_3$, Nature Physics 2013, 9, 410-414.

**(22)** Edmonds, K.W.; Farley, N.R.S.; Campion, R.P.; Foxon, C.T.; Gallagher, B.L.; Johal, T.K.; van der Laan, G.; MacKenzie, M.; Chapman, J.N.; Arenholz, E. ; Surface effects in Mn $L_{2,3}$ x-ray absorption spectra from (Ga,Mn)As. *Appl. Phys. Lett.* **2004**, *84*, 4065-4067.

**(23)** Maccherozzi, F.; Sperl, M.; Panaccione, G.; J. Minár, J.; Polesya, S.; H. Ebert, H.; Wurstbauer, U.; M. Hochstrasser, M.; Rossi, G.; Woltersdorf, G.; et al. Evidence for a Magnetic Proximity Effect up to Room Temperature at Fe/(GaMn)As Interfaces. *Phys. Rev. Lett*. **2008,** *101***,** 267201.

Entry for the Table of Contents

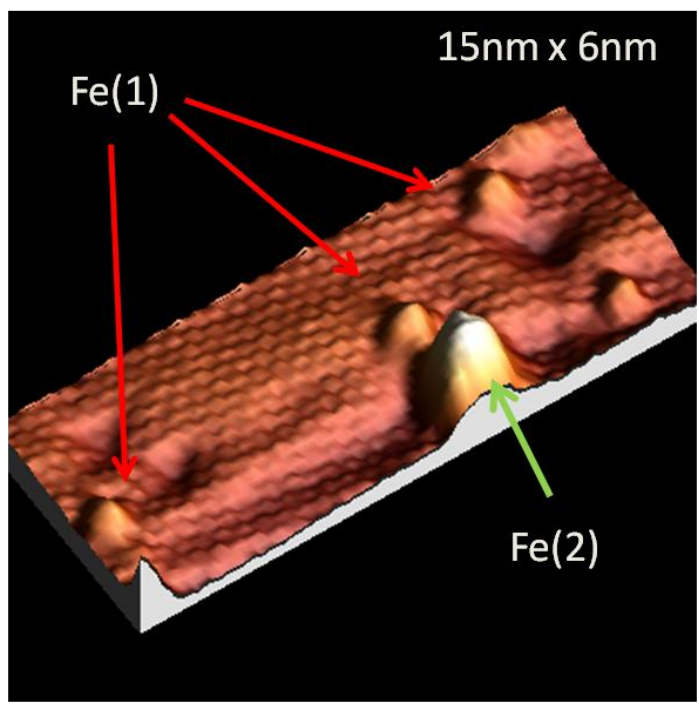

# Supporting Information

**Supporting Information : Contents**

1. Experimental and theoretical details

2. Self doping effect in as-cleaved $Bi_{2-x}Mn_xTe_3$ and $Bi_2Te_3$ crystals

3. Comparison of XPS and HAXPES Ti *4d*, Bi *5d* and Bi *4f* core level

4. Additional calculation details

## 1. Experimental and theoretical details

**Sample preparation and characterization:** High purity elemental Bi (99.999%), Mn (99.99%), and Te (99.999%) were used for the $Bi_{2-x}Mn_xTe_3$ crystal growth, employing a nominal x value from 0.03 to 0.1. A modified Bridgeman crystal growth method was employed. The crystal growth for $Bi_{2-x}Mn_xTe_3$ involved cooling from 950 to 550 °C over a period of 24 h and then annealing at 550 °C for 3 days; silver-colored single crystals were obtained. The crystals were confirmed to be single phase and identified as having the rhombohedral $Bi_2Te_3$ crystal structure by X-ray diffraction. Crystals were p-type with a carrier concentration between $10^{19}$ and $10^{20}$ carrier/cm$^3$. The Mn concentration in the bulk crystals was determined by the ICP-AES method and was very close to the nominal compositions. Details of the structural and magnetic characterization as a function of temperature, magnetic field and doping can be found in Refs. 3 and 4 of the main manuscript text. Clean surfaces of $Bi_2Te_3$ and $Bi_{1.91}Mn_{0.09}Te_3$ were obtained by cleaving their respective crystals *in-situ* (pressure better than 5 X $10^{-11}$ mbar) at T= 22 K (soft X-ray XPS and XAS), at T= 6K (ARPES) and at room temperature (HAXPES). Fe was deposited on $Bi_2Te_3$ by the e-beam evaporation technique with the substrate crystal kept at room temperature. The deposition rate was monitored using a quartz microbalance and was calibrated to be ≤ 0.1 Å/min. During the whole process, the vacuum inside the chamber was better than $2 \times 10^{-10}$ mbar.

**STM measurements:** STM measurements were performed using a home-built room temperature UHV STM. A chemically etched W wire was used for a tip. The STM topographic images were acquired with a constant current mode. The sample bias voltage and the tunneling current for each image are indicated in the figure captions. The samples were then transferred in UHV conditions from the STM/Fe growth chamber to the end station for the XAS/XPS measurements and vice-versa[1].

**X-ray techniques and experiments:** We acquired XAS spectra at the APE beamline of the Elettra Synchrotron (Trieste, Italy) in the temperature range of 20-300 K and in a base vacuum $< 2 \times 10^{-10}$ mbar.[1] X-ray Magnetic circular Dichroism (XMCD) was used to check the ferromagnetic properties of Fe and Mn, similarly to Ref 9. All XAS and XMCD measurements were obtained by measuring the sample drain current - Total Electron Yield (TEY) - with an energy resolution of 150 meV and a circular polarization rate of 70%. TEY mode was used to enhance sensitivity of the near surface-interface region (TEY depth sensitivity is in the range of 5 nm). The photon incidence angle was kept fixed at 45° from the sample surface normal. Beam spot size on the sample surface was 200x300 μm². All spectra were normalized by the intensity of the incident beam, which is given by the TEY of a gold mesh. The comparison between experimental XAS lineshapes of Mn $L_{2,3}$ was obtained after background subtraction. The raw Mn $L_{2,3}$ XAS spectra displayed are first normalized to the incoming photon flux measured at the same time on the reference mesh. A straight baseline fitting the slope of the raw TEY signal and matching the intensity of the $L_3$ pre-edge region has been subtracted to the data.

**XPS-HAXPES.** XPS experiments have been performed at APE beamline by means of an Omicron EA-125 photoelectron analyzer in the same analysis chamber of the XAS experiments.[1] Both XPS and XAS/XMCD measurements were possible at identical sample positions and with identical spot size. XPS data have been acquired at hν =455 eV in normal emission geometry using linearly polarized X-rays, at 30 K for freshly cleaved $Bi_2Te_3$, $Bi_{1.91}Mn_{0.09}Te_3$ surfaces, and at both room temperature and 30 K after Fe deposition. No difference has been observed with temperature. The overall energy resolution (photons + analyzer) was about 250 meV. No trace of oxidation and/or contamination, by regularly checking the presence of Carbon *1s* and Oxygen *1s* peaks in survey spectra, was detected over completion of one set of measurements, after which a new fresh cleaved surface was used. We cannot exclude a small oxidation of Bi and Te under Fe evaporation, however. Following our fitting procedures, the presence of oxygen-related extra peaks hidden in the large Te *4d* and Bi *5d* components is limited to a few percent (<3%).

HAXPES measurements were carried out using linearly horizontal polarized light at the SOLEIL synchrotron (Paris-France), using the undulator beamline GALAXIES. The X-ray incidence angle as measured from the sample surface was fixed at 2 degrees. Samples have been cleaved before insertion in vacuum, and measured at room temperature. The overall energy resolution (monochromator + analyzer) was set to 250 meV (2.5 keV), and 0.8 eV (8 keV). For both HAXPES and XPS Binding energy and energy resolution has been estimated by means of a reference polycrystalline gold foil in electrical and thermal contact with the sample, measuring Au 4f core level and Fermi energy.

**ARPES.** ARPES experiments were performed on samples cleaved at T =6 K, in a vacuum better than $5 \times 10^{-11}$ torr, at the Quantum Materials Laboratory of the AMPEL Lab (UBC, Vancouver, Canada), with 21.2 eV linearly polarized photons on a dedicated spectrometer equipped with a SPECS Phoibos 150 hemispherical analyzer and UVS300 monochromatized gas discharge lamp. Energy and angular resolution were set to 10 meV and $\pm 0.1^0$, respectively. The orientation of the samples was checked by Laue diffraction peaks and samples were aligned with the Γ to K direction parallel to the entrance slit of the analyzer.

**Calculation details:** The Hamiltonian for the initial and final state in the photoemission process were calculated using Cowan's code[2], including spin-orbit and multiplet structure but neglecting band structure dispersion. In the Mn atom the $3d^5$ configuration has the lowest energy but is strongly mixed with the $3d^6$ configuration, which is at slightly higher energy. The wave functions are given as a coherent sum over $d^n$ states with different electron number, $n$, embedded in a medium which accepts electrons at the cost of a charge-transfer energy $\Delta$. Wave functions were calculated in intermediate coupling using the atomic Hartree-Fock approximation with relativistic corrections and scaling the Slater parameters to 70% to account for intra-atomic correlation effects. The initial state was modelled as 67.8% $d^5$, 29.1% $d^6$, and 3.1% $d^7$. The resulting line spectrum was broadened by a Lorentzian of $\Gamma = 0.5$ eV and a Gaussian of $\sigma = 0.4$ eV to account for life-time, instrumental, and inhomogeneous broadening.

## 2  Self doping effect in as-cleaved $Bi_{2-x}Mn_xTe_3$ and $Bi_2Te_3$ crystals

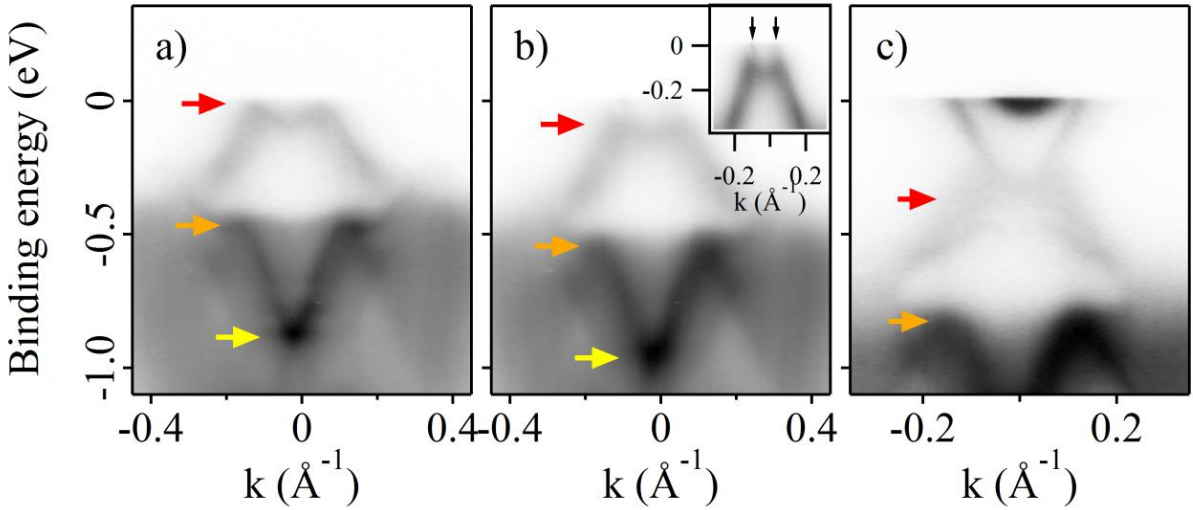

**Fig. S2**. As written in the text of the main paper, the Dirac cone in as-cleaved samples is located above $E_F$ due to the intrinsic hole doping, a feature well in agreement with previous reports (Refs. 17 and 18 in the main manuscript text). This p-type character of $Bi_2Te_3$ samples is in contrast to the n-type $Bi_2Se_3$ samples, which always have a Fermi level crossing the bulk conduction bands. For pure $Bi_2Te_3$, the characteristic Dirac cone is observed only upon electron doping; this occurs spontaneously, as a function of time after exposing the freshly cleaved surface, or by surface doping, e.g. via K deposition. 12 hours

later the Dirac point is 160 meV below $E_F$. Because of the surface instability, it is difficult to identify the Mn impurities as hole dopants simply by comparing the Fermi level positions. The figure represents the band structure along Γ-K for (a) as cleaved Mn x=0.04 sample and (b) the same sample 10 hours after the cleave; the arrows point to the most prominent band structure features and emphasize the rigid energy shift (75 meV towards higher binding energies) as a function of time; the inset emphasizes the near-$E_F$ area of aged sample where Dirac cone starts to be visible (black arrows); (c) the same sample after K-doping: the band structure shifts by additional 250 meV and the Dirac point falls at ~380 meV. These data are relative to 4%Mn, but the effect is generic for $Bi_2Te_3$.

## 3   Comparison of XPS and HAXPES for Ti *4d*, Bi *5d* and Bi *4f* core level

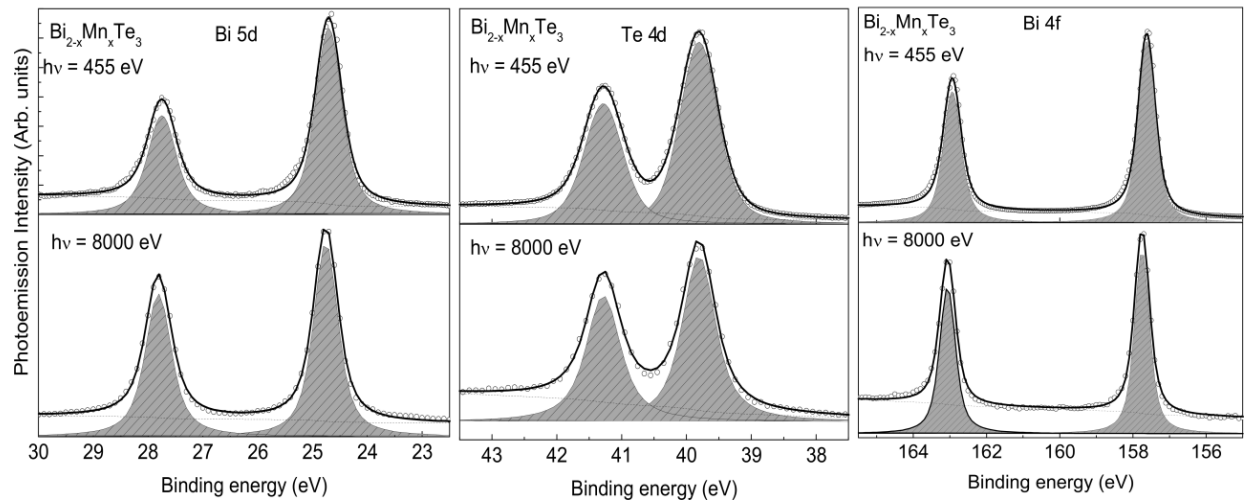

**Fig. S3**. We have also performed a fitting analysis on Bi *4f*, Bi *5d* and Te *4d* shallow core level, for both pure $Bi_2Te_3$, and Mn-doped $Bi_{1.91}Mn_{0.09}Te_3$, comparing their lineshapes as measured at 455 eV and in the HAXPES regime (8 keV), similarly to the comparison presented in figure 4 only for soft X-ray. No difference in lineshapes, within the limit of the different energy resolution (0.25 eV at 455 eV and 0.8 eV at 8 keV) has been found, reinforcing the result of a diluted character of Mn. Black curves are results of fitting, superimposed at raw data (symbols). Grey peaks are fitting results of individual peaks, shaded curve are the background of the peak, that has been subtracted from raw data.

## 4  Additional calculation details

The Mn 1s photoemission spectrum corresponding to the transition $3d^n \rightarrow 1s^1 3d^n \varepsilon$ can be calculated using an Anderson impurity model, taking into account configuration interaction in the initial and final state.[2] The wave functions are given as a coherent sum over $d^n$ states with different electron number, $n$, embedded in a medium which accepts electrons at the cost of a charge-transfer energy $\Delta$. For sake of explanation for the Mn core-level photoemission it is sufficient to consider only the $d^5$ and $d^6$ states. In the Mn atom the $3d^5$ configuration has the lowest energy but is strongly mixed with the $3d^6$ configuration, which is at slightly higher energy. The case of the Mn 2p photoemission from (Ga,Mn)As has been described in Ref. 4. While in the ground state the $3d^5$ level has lower energy than the $3d^6$ level, in the final state the $2p^5 3d^6$ is pulled down in energy below the $2p^5 3d^5$ due to the strong core-valence Coulomb interaction, $Q = 5$ eV. This gives a Mn 2p PE spectrum with a low energy peak dominated by the screened $2p^5 3d^6$ final-state configuration while the peak at higher binding energy has mainly $2p^5 3d^5$ character. An important difference between a 1s and 2p core hole is the degree of localization as well as screening, which affects the 3d valence states. The core-valence Coulomb interaction, $Q(1s,3d)$, and the intra-atomic Coulomb interaction, $U(3d,3d)$, will be smaller in the presence of a 1s core hole than for a 2p hole.[5] However, $Q(1s,3d)$ is still large enough to pull down the $1s^1 3d^6$ state in energy below the $1s^1 3d^5$ state. Such a level reversal leads to the appearance of a strong satellite feature.[6]